\documentclass[aps,groupedaddress,twocolumn,showpacs,letterpaper]{revtex4}
\usepackage{graphicx} % standard LaTeX graphics tool
\usepackage{amsmath} 
\begin{document}
\title{Engineering single atom and single photon sources from entangled atomic ensembles }

\author{ M. Saffman and T. G. Walker}
\affiliation{
Department of Physics,
University of Wisconsin, 1150 University Avenue,  Madison, Wisconsin 53706
}
 \date{\today}

\begin{abstract}
We discuss the application of dipole blockade techniques for
 the preparation of single atom and single photon sources. A deterministic protocol is given for loading a single atom in an optical trap as well as ejecting a controlled number of atoms in a desired direction. A single photon source with an optically controlled beam-like emission pattern is described.    
\end{abstract}

\pacs{32.80.-t,32.80.Qk,03.67.Dd}
\maketitle

Quantum information science builds on the laws of quantum mechanics to transmit, store, and process information in new and powerful ways. Advances in this field rely on our ability to manipulate coherently isolated quantum objects while eliminating incoherent interactions with the surrounding environment. It was proposed several decades ago that the availability of single photon sources would enable secure transmission of information without risk of eavesdropping\cite{ref.bb84}. Single photon sources have been proposed and realized  using a number of different approaches\cite{ref.spsources}, and 
demonstrated for information transmission\cite{ref.qibook}. 
 Another example of a physical embodiment of quantum information is provided by  a single neutral atom, the internal states of which can be used to represent a 
quantum bit or ``qubit". Placing individual atoms on a lattice defined by optical beams and allowing them to interact is one of the approaches currently being explored for constructing a 
quantum computer \cite{ref.atomcomputer}. 

In this letter we describe some new approaches to engineering single atom and single photon sources for use in quantum information science. 
These ideas utilize the proposal of  Lukin et al.\cite{ref.blockade2}  
for entanglement of mesoscopic atomic ensembles using a  dipole-blockade mechanism that is mediated by dipole-dipole interactions of highly excited Rydberg atoms. Lukin et al. emphasized the application of mesoscopic many-atom qubits for quantum logic. 
Here we discuss how to combine many-atom entanglement with laser cooling and trapping techniques for loading a single atom into an optical trap, as well as creating single atom and single photon sources. We show that both   atoms and photons can be extracted with  well defined propagation directions.

\begin{figure}[!t]
\begin{minipage}[c]{7.5cm}
\includegraphics[width=7.5cm]{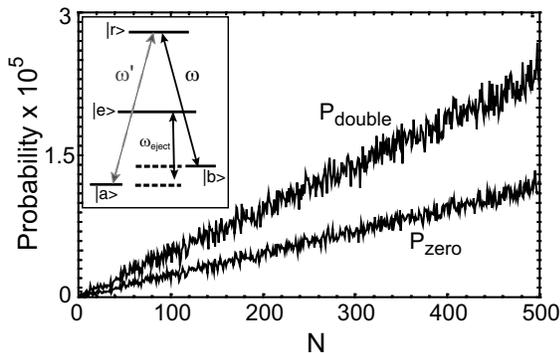}
\caption{Probability of non-excited and doubly excited states after a Rydberg pulse as a function of the number of atoms. The inset shows the atomic level scheme.  }
\label{fig.p1p2calc}
\end{minipage}
\end{figure}

Consider a collection of $N$ atoms at positions ${\bf r}_j$ each with  non-degenerate ground states $|a\rangle$, $|b\rangle$, intermediate states $|e\rangle$ and highly  excited Rydberg states $|r\rangle$ as shown in the inset of Fig. \ref{fig.p1p2calc}.  We envision that the atoms have been laser cooled to $\mu$K temperatures, and confined to a volume of a few $\mu\rm m^3$ defined by an optical trap created by far off-resonance optical beams (FORT trap) using standard techniques\cite{ref.metcalfbook}.  
Following the theory of Lukin, et al.\cite{ref.blockade2} we define  the following collective atomic states:  the ground state  $|g\rangle=\prod_j |b_j\rangle=|b_1b_2...b_N\rangle,$ the singly excited state 
$|r_j\rangle=|b_1... r_j ...b_N\rangle,$ and the doubly excited state 
$|r_j r_k\rangle= |b_1... r_j .... r_k ...b_N\rangle.$ The interaction  Hamiltonian governing the coupling of levels $|b\rangle$ and $|r\rangle$ assuming zero atom-field detuning  is taken to be $\hat V = \hat V_d + \hat V_{dd}$ where 
$2 V_d/\hbar =  
  \sum_j \Omega_j  |r_j\rangle\langle g| 
+  \sum_{j,k>j}\Omega_k  |r_j r_k \rangle\langle r_j|
 + {\rm  H.c.}  $ 
describes the electric-dipole atom-field coupling in the rotating wave approximation, and $V_{dd}/\hbar=\sum_{j,k>j} \Delta_{jk}  |r_j r_k \rangle\langle r_j r_k|$ describes the dipole-dipole interaction of two excited atoms. Here  $\Omega_j=-\langle r_j|
 \hat d |b_j\rangle {\mathcal E({\bf r}_j})/\hbar,$ where $\hat d$ is the dipole moment operator, and  the 
position dependent optical field 
is $E({\bf r}_j) =({\mathcal E} ({\bf r}_j) /2)
e^{-\imath\omega t} + c.c.$ . The dipole-dipole shift in the case of dipole moments aligned parallel to the vector separating the atoms is given by $\Delta_{jk} = - f(n) e^2 a_0^2 / |{\bf r}_j - {\bf r}_k|^3$ with $e$ the electronic charge and $a_0$ the Bohr radius. The numerical factor $f(n)$ can be found from a quantum defect theory calculation of the Rydberg Stark  map. In the presence of a hybridizing dc field  $f(n)\sim n^6$ for a level with principal quantum number $n$\cite{ref.gallagherbook}.

When $\Delta_{jk}\gg|\Omega_j|$ we can safely neglect triply excited states and higher. An arbitrary $N$ atom  state vector can be written in this limit  as $|\psi(t)\rangle= c_g(t) |g\rangle + \sum_j \tilde c_j(t) |r_j\rangle + \sum_{j,k>j}  \tilde c_{jk}(t)|r_jr_k\rangle$.
We can obtain analytical estimates of the dynamics by considering 
the atomic evolution due to a field that is pulsed on to excite a transition to a Rydberg state and has a  spatially uniform intensity  such that $\Omega_j=\Omega e^{\imath \phi_j}.$ We neglect the atomic motion during a pulse so that $\phi_j$ is taken as constant.  Incorporating the phases $\phi_j={\bf k}\cdot {\bf r}_j$ for a traveling wave with wavevector $\bf k$ into the amplitudes $ c_j = \tilde c_j e^{-\imath \phi_j}$ and $ c_{jk}=\tilde c_{jk} e^{-\imath (\phi_j+\phi_k)},$ assuming all coefficients $ c_j$  equal, and adiabatically eliminating the doubly excited state,   the 
normalized  singly excited symmetric state can be written as 
\begin{equation}
|s\rangle = \frac{1}{\sqrt N}\sum_j  e^{\imath \phi_j} |r_j\rangle 
\label{eq.single}
\end{equation}
and has amplitude   $c_s= \sqrt N  c_j.$ The Schr\"odinger equation  
 results in 
\begin{subequations}
\begin{eqnarray}
\dot c_g &=&  -i\frac{\sqrt N \Omega^*}{2}  c_s, \\
\dot{  c}_s &=&   -i\frac{\sqrt N \Omega}{2}c_g +i\frac{(N-1)|\Omega|^2}{2 \bar\Delta_{dd}}c_s,
\end{eqnarray}
\label{eq.sym}
\end{subequations}

\noindent
with $\bar\Delta_{dd}=(N(N-1)/2)\left[\sum_{j,k > j} 1/\Delta_{jk}\right]^{-1}$. Solving Eqs. (\ref{eq.sym}) with the initial condition 
 $c_g(0)=1$ (this is readily achieved with optical pumping techniques) we find
 $|c_s(t)|^2= l^{-1}\sin^2[\frac{\sqrt {Nl} |\Omega|}{2} ~t]$  
with $l= 1+ (N-1)^2 |\Omega|^2/4N \bar \Delta_{dd}^2.$ Thus at time $t=\pi/(\sqrt{Nl} |\Omega|)$ we have rotated the ground state 
to the singly excited symmetric state  $|s\rangle$ with probability $P_{\rm single}=1/l$ and since Eqs. (\ref{eq.sym}) conserve probability 
$|c_g(t)|^2=P_{\rm zero}=1-P_{\rm single}.$
The unwanted  leakage into the doubly excited states at  the end of the $\pi$ pulse  is found by summing over the doubly excited probabilities resulting in 
\begin{equation}
P_{\rm double}\sim\sum_{j,k>j} ~ | c_{jk}|^2\sim \frac{N-1}{2l}\frac{|\Omega|^2}{\bar \Delta_{dd}^2}.
\label{eq.p2}
\end{equation}
%
%%%%%%%%%%%%%%%%%%%%%%%%%%%%%%%%%%%%%%%%%%%%%%%%%%%%%

We can estimate the fidelity of preparation of a singly excited state starting with $N$ trapped atoms using physical parameters of $^{87}$Rb as an example.  Assume that the atoms are randomly distributed in a sphere of  $5~\mu\rm m$ diameter and that we are using the $n=50$ Rydberg levels. A quantum defect calculation shows that a pair of atoms with the maximum separation of $5~\mu\rm m$ subject to a hybridizing dc field experience a dipole-dipole shift of 
$\Delta/2\pi \sim 100$ MHz. Using $|\Omega|/2\pi=1$ MHz we see, as shown in Fig. \ref{fig.p1p2calc}, that the probability of creating a state with zero or two excitations  grows linearly with $N$ (the average interatomic spacing and $\bar\Delta_{dd}$ stay roughly constant as atoms are added) and remains less than $3\times 10^{-5}$ for up to $N=500$ atoms.

After creating  the singly excited state we   apply a $\pi$ pulse at $\omega'$ to transfer the single-atom excitation to
the lower ground state $|a\rangle$. The collective atomic state after this pulse sequence  is proportional to $(1/\sqrt N) \sum_{j=1}^N e^{\imath \phi_j}|b_1...a_j....b_N\rangle$, which has the same functional form as Eq. (1). 
It can be shown that corrections to the $N$ atom results due to spontaneous emission  are $O(N \gamma_{R}/\bar\Delta_{dd})$ with $\gamma_R$ the Rydberg level spontaneous decay rate. The correction is negligible for experimental conditions of interest with up to several thousand atoms. 
It should be noted that it is essential for high fidelity preparation that the pulses be applied sequentially. Applying both $\omega$ and $\omega'$ simultaneously leads to large amplitude multiply excited leakage from $|b\rangle$ to $|a\rangle$
due to multiphoton Raman processes.  

\begin{figure}[!b]
\begin{minipage}[c]{8.5cm}
\includegraphics[width=8.5cm]{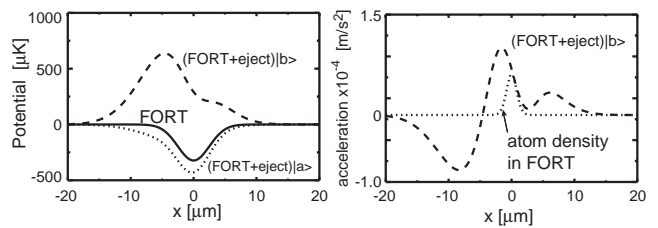}
\caption{Potentials(left) and acceleration(right) for atom ejection with the eject beam centered $3~\mu\rm m$ to the left of the FORT beam. Calculations for $^{87}$Rb with  $w_{\rm FORT}=5~\mu\rm m,$ $P_{\rm FORT}=100~\rm mW$, $\lambda_{\rm FORT}=1.06~\mu\rm m,$ $w_{\rm eject}=10~\mu\rm m,$ $P_{\rm eject}=9~\mu\rm W$, $(\omega_{\rm eject}-\omega_{rb})/2\pi 
= 1 \rm ~ GHz,$ and $T_{\rm atomic}=30~\mu\rm K.$ }
\label{fig.ejectforce}
\end{minipage}
\end{figure}

One application of this many-atom entangled state is to load an array of optical traps  with a single atom in each one for use as a quantum computer\cite{ref.atomcomputer}.   In that application the dipole blockade mechanism can also be used between adjacent qubit sites to implement two-qubit logic gates\cite{ref.blockade1}. 
With $N$ atoms in a given lattice site we use the above procedure to prepare the state $|a\rangle.$
To eject the $N-1$ atoms remaining in state $|b\rangle$ we apply a beam that causes  strong mechanical forces on $|b\rangle$ but only weak forces on $|a\rangle.$ One possibility is to use a beam with a waist a few times larger than the FORT beam waist $w_{\rm FORT}$ 
and with a frequency $\omega_{\rm eject}$ that is tuned to the red of $\omega_{ea}$ and to the blue of $\omega_{eb}$ as shown in Fig. \ref{fig.p1p2calc}.  This gives repulsive gradient forces on $|b\rangle$ and attractive forces on $|a\rangle$ as shown numerically  in Fig. \ref{fig.ejectforce}. In the left hand plot we see that the combined potential from the FORT beam and the eject beam is strongly repulsive for $|b\rangle,$ but attractive with only a small shift in the position of the minimum for $|a\rangle.$  The net acceleration of atoms in state $|b\rangle$ due to the FORT and eject beams  is positive causing motion to the right over the entire region occupied by the atom cloud. 
After a characteristic time $t_1$ determined by $\frac{1}{2} a t_1^2=w_{\rm FORT}$ with $a=F/m$ the net acceleration, atoms in state $|b\rangle$ will be swept free of the FORT potential,leaving a single atom in state $|a\rangle$ behind. 
Using the parameters in Fig. \ref{fig.ejectforce} we get  $t_1\sim 40 ~\mu\rm s.$
 Note that since the indistinguishable atoms are actually in an entangled superposition state we cannot actually speak of 1 atom left behind in the trap and $N-1$ ejected until a measurement or decoherence has caused projection of the wavefunction. Finally it should be emphasized that although several experiments have achieved single atom loading in optical traps, they have relied on stochastic loading into  extremely small volumes\cite{ref.1atomtrap}. High fidelity single atom loading  is essential for filling a large number of traps in a neutral atom 
quantum processor\cite{ref.mott}.

As suggested by Lukin et al.\cite{ref.blockade2} a deterministic beam of atoms can also be generated using the entangled ensemble. 
We use a  
protocol very similar to that described above for single atom loading but now
  start by optically pumping the trapped atoms to state $|a\rangle.$ 
A sequence of $\pi$ pulses on $\omega'$ and $\omega$ will transfer a 
single excitation to $|b\rangle$ which can be ejected from the trap as described above. To obtain multiple atom pulses, each containing $m$ atoms, we first go  through $m$ dipole blockade cycles to create an entangled state with $m$ units of excitation in $|b\rangle$\cite{ref.blockade2}. We then eject the $m$ atoms by applying 
$\omega_{\rm eject},$ recreate the $m-$times excited state, eject the atoms, and so on. We can do this roughly $N/m$ times in a deterministic fashion 
before having to reload the FORT with atoms. Assuming single atom Rabi frequencies of 1 MHz the time it takes to create an $m-$times excited state is roughly $m/\sqrt N$, since the effective Rabi frequencies scale as $1/\sqrt N.$ We see that for pulses with up to about a hundred atoms the repetition rate will be limited by the ejection time which we have estimated as $t_1\sim 40 ~\mu\rm s.$  While the parameters can  be chosen for faster or slower operation, kHz atomic pulse rates are certainly accessible. 

The direction of the ejected atoms will be determined by the positioning of the eject beam with respect to the FORT, which provides a means of 
 scanning the ejected atoms. The transverse spread of the atomic beam will be determined by gradients in the eject beam, as well as fluctuations due to spontaneous emission.  
 For the 1 GHz detuning used in the numerical example of Fig. \ref{fig.ejectforce} the number of photons scattered from atoms in $|b\rangle$ during a $t_1\sim 40~\mu\rm s$ eject pulse is $n_{\rm scat}\sim 21.$  (The corresponding number of photons scattered from atoms in $|a\rangle$ for which the detuning is -5.8 GHz is only about 0.6.)
This results in an rms momentum transfer due to spontaneous emission  that is about a tenth of the coherent impulse after an eject time $t_1.$ 
The emitted beam will therefore be well collimated.  
It is also feasible to move atoms a controlled distance using traveling 
dipole force fields as demonstrated recently in \cite{ref.meschededelivery}.
The  flexibility of optical control of the beam direction opens the potential of many applications in areas that include optical lattice neutral atom quantum computers, atomic interferometers, precision low-level current sources, and also nanofabrication tasks at the single atom level.

\begin{figure*}[!]
%%\begin{minipage}[c]{16cm}
\includegraphics[width=16.cm]{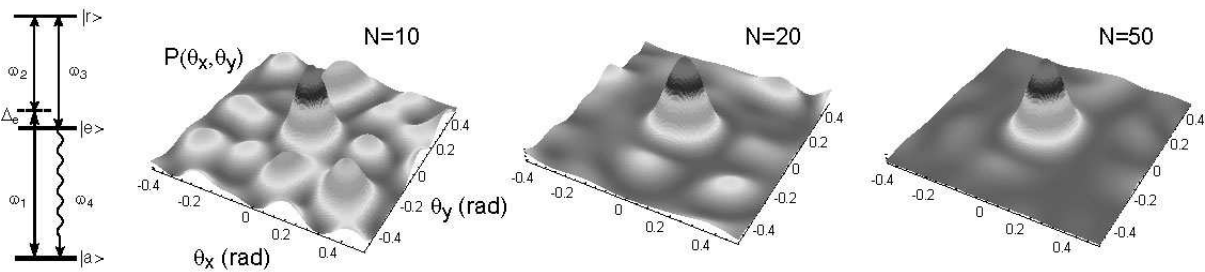}
\caption{Fields used for state preparation and angular emission probabilities (arb. units on linear scale) at $\lambda_4=0.78~\mu\rm m$ for $N=10,20,$ and $50$ atoms randomly 
positioned within a $5~\mu$m diameter sphere. }
\label{fig.angdist}
%%\end{minipage}
\vspace{-.5cm}
\end{figure*}

We turn now to the creation of a phased array single photon source with a diffraction-limited emission pattern.
When the optical fields propagate through the sample, atoms will be excited with position dependent relative phases as in Eq. (\ref{eq.single}). This results 
in an entangled state with a phase structure that mimics the phase of the exciting beam and can be used to create a single photon source with well defined directionality.  While previous work has achieved single photon sources with a controlled emission direction by coupling to 
microcavities\cite{ref.spdirectional}
our approach results in a source emission pattern that is reconfigurable and is defined by the structure of the preparation optical fields.  
Referring to  Fig.~\ref{fig.angdist}, fields at 
$\omega_1$ and $\omega_2$ with Rabi frequencies $\Omega_1$ and $\Omega_2$ drive a 
two-photon transition
$|a\rangle \rightarrow |e\rangle \rightarrow |r\rangle$ to the Rydberg level $|r\rangle$ in an $N$-atom
ensemble.  The effective Rabi frequency for the two photon process acting on atom $j$ is 
$|\Omega| e^{i\phi_1+i\phi_2}=
(|\Omega_1|e^{i\phi_1({\bf r}_j)}|\Omega_2|e^{i\phi_2({\bf r}_j)})/
\Delta_e,$ 
 where $\phi_1$ and $\phi_2$ are the phases of fields 
$\omega_1, ~\omega_2$ at the atomic position ${\bf r}_j$  and $\Delta_e=\omega_1-\omega_{ea}.$
As long as $P_{\rm double}$ given by Eq. (\ref{eq.p2}) is small  only
transitions to states with a single excited atom are energetically 
allowed and we have an effective dipole blockade.
Under these conditions an ensemble of atoms in $|a\rangle$ subjected to a   $\pi$-pulse applied on 
$|a\rangle\rightarrow |r\rangle$ 
produces the entangled symmetric superposition
state
$|\psi\rangle=\frac{-i}{\sqrt{N}}\sum_je^{\imath(\phi_{1j}+\phi_{2j})}|r_j\rangle.
$
 The phases $\phi_{mj}={\bf k}_m\cdot {\bf r}_j$ are 
simply the phase of the $m^{\rm th}$ laser field  ($\sim e^{\imath({\bf k}_m\cdot{\bf r} - \omega_mt)})$ at the
position
${\bf r}_j$ of the
$j^{\rm th}$ atom.

We now apply a $\pi$-pulse with $\omega_3$ tuned to the 
$|r\rangle \rightarrow |e\rangle$ single-photon transition.  The wavefunction 
is then  transformed into
\begin{equation}
|\psi\rangle=\frac{-1}{\sqrt{N}}\sum_je^{\imath(\phi_{1j}+\phi_{2j}-\phi_{3j})}|e_j\rangle.
\end{equation}

This state will radiate into a variety of modes with all the atoms in 
state $|a\rangle$ and a single photon propagating
in  direction ${\bf k_4}$.  The amplitude for emission into state 
$|g,1_{{\bf k}_4}\rangle=|a_1...a_j...a_N,1_{{\bf k}_4} \rangle$ is
proportional to 
%
%%%ie\sqrt{\frac{2\pi\hbar\omega_4}{V} }
%
\begin{eqnarray}
&&\langle g,1_{{\bf k}_4}|({\bf e_{k_4}}\cdot {\bf \hat d})
{\hat a}_{{\bf k}_4}^\dagger e^{\imath{{\bf k}_4\cdot {\bf r}}}|\psi,0_{{\bf k}_4}\rangle\nonumber
\\&=&
-
\sum_j \frac{\langle g|{\bf e_{k_4}}\cdot 
{\bf \hat d}|e_j\rangle e^{\imath({\bf k_4-k_1-k_2+k_3}) \cdot{\bf r}_j}}{\sqrt{N}}
\end{eqnarray}
where ${\bf e_{k_4}}$ is the polarization of the emitted photon 
 and
${\hat a}_{{\bf k}_4}^\dagger$ is the creation operator for a photon in mode ${\bf k}_4$. 
Assuming an isotropic angular distribution
from the atomic matrix element gives for the angular distribution
\begin{equation}
P({{\bf k}_4})\propto \frac{1}{N}\left|\sum_j e^{i({\bf k_4-k_1-k_2+k_3)\cdot{\bf r}_j}}\right|^2.
\label{eq.pofk}
\end{equation}
Note that the emission is highly directional; in the direction ${\bf 
k_4=k_1+k_2-k_3}$
the phases are zero for each
atom giving
$P({{\bf k}_4})=N$
while for other directions the phase factors become random and 
$P({{\bf k}_4})\sim 1$.   A
calculation of the angular distribution, assuming an isotropic atomic 
matrix element, is shown in
Fig.~\ref{fig.angdist}. The width of the central emission peak is
determined by diffraction. For $N=50$ the numerical value for the FWHM agrees to better than 10\% with the diffraction estimate of  $\lambda/D$, where $D$ 
is the diameter of the phased atom cloud.

\begin{figure}[!b]
\begin{minipage}[c]{8cm}
\includegraphics[width=6.cm]{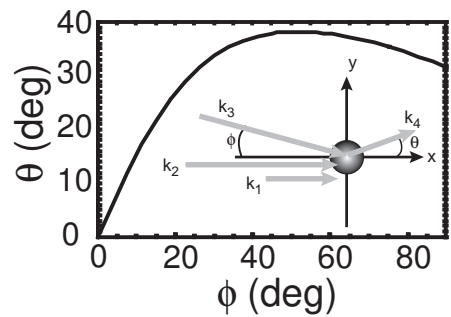}
\caption{Angular direction $\theta$ of the photon emitted along ${\bf k}_4$ for collinear Rydberg excitation beams ${\bf k}_1,{\bf k}_2$ and ${\bf k}_3$ tilted by an angle $\phi.$ }
\label{fig.sps}
\end{minipage}
\end{figure}

It is evident from Eq. (\ref{eq.pofk}) that the angular pattern of the emitted single photon beam bears  a close analogy with four-wave mixing. If we arrange for ${\bf k}_2=-{\bf k}_1$ the photon will be emitted in a phase conjugate mode with ${\bf k}_4=-{\bf k}_3.$ For a single photon source it is more convenient to have the emitted photon angularly separated from the other beams as shown in Fig. \ref{fig.sps}. By setting the waist size of the ${\bf k}_1, {\bf k}_2, {\bf k}_3$ beams to be much larger than the size of the 
atomic distribution all atoms  will to a good approximation be driven by  the same optical intensities so that the accuracy of the preparation Rabi pulses will be high. The angular divergence of the emitted pulse will  be determined by the size of the atomic cloud, as seen  in Fig. \ref{fig.angdist}.  

Several factors can contribute to angular broadening of the emitted mode. 
The  atomic motion  during  a  $3~\mu$s sequence of preparation pulses, accounting for the characteristic speed  of $30~\mu$K $^{87}$Rb atoms, is $\Delta x\sim 0.15~\mu$m, which is about $1/5$ of the emission wavelength. The contribution of this motional dephasing to broadening of the emission mode is a subject of current study.   Imperfect dipole blockade leading to doubly excited states 
can also degrade the single photon fidelity. We note that this will be strongly suppressed by angular selection since 
the contribution to emission along ${\bf k}_4$
due to a  doubly excited Rydberg state channel  is proportional to 
$(1/N)|\sum_j e^{\imath(k_4-2(k_1+k_2)+k_3)}|^2$ which is not phase matched along the direction $\theta$ selected in Fig. \ref{fig.sps}. 
Finally we note that by employing an excited level $|e\rangle$ that is not dipole coupled to the ground state it will be possible to create entangled two-photon pairs with similar beam-like emission properties.

This work is supported by the NSF and NASA.
M. S. is an A. P. Sloan Foundation fellow.

\end{document}